\documentclass[journal]{IEEEtran}

\ifCLASSINFOpdf
\else
   \usepackage[dvips]{graphicx}
\fi

\usepackage{graphicx}
\usepackage{textcomp}
\usepackage{stfloats}
\usepackage{url}
\usepackage{verbatim}
\usepackage{bm}
\usepackage{cite}
\usepackage{amsmath,amsfonts,amssymb}
\usepackage{algorithmic,algorithm}
\usepackage{array}
\usepackage{subfig}
\usepackage{float}
\allowdisplaybreaks[4]

\begin{document}

\title{RFDT-Channel: RGB-LiDAR-Based RF Digital Twin Scene Construction for 28 GHz Indoor Ray-Tracing Channel Simulation}

\author{Chengyang Yao, Cunhua Pan, Jiaming Zeng, Yuquan Sun, Haoyang Weng, Haojian Wang, Hong Ren,  Jiangzhou Wang
\vspace{-0.5cm}
\thanks{Chengyang Yao, Cunhua Pan, Jiaming Zeng, Yuquan Sun, Haoyang Weng, Haojian Wang, Hong Ren, and Jiangzhou Wang are with the School of Information Science and Engineering, Southeast University, Nanjing 210096, China. Corresponding author: Cunhua Pan. Contact e-mail: young929727@163.com.}}

\maketitle

\begin{abstract}
Real-scene indoor millimeter-wave simulation requires efficient modeling of radio frequency (RF)-computable geometry and electromagnetic material properties. To address the low efficiency of manual scene modeling, the limited RF adaptability of visually reconstructed meshes, and the lack of material binding in 28~GHz ray-tracing simulation, RFDT-Channel is developed as an RF digital twin scene construction workflow based on red-green-blue (RGB) images and light detection and ranging (LiDAR) point clouds. Indoor videos and point clouds are collected by a Jetson Orin platform with LiDAR and GMSL cameras. An initial triangular mesh is generated through COLMAP, 3D Gaussian Splatting (3DGS), and SuGaR. The LiDAR point cloud then provides geometric and scale references for RF-oriented regularization in Blender, including alignment, wall solidification, door/window opening construction, and topology repair. OpenScene semantic segmentation maps major indoor structures to concrete, glass, wood, and metal materials, and Sionna RT performs 28~GHz ray tracing. Under a fixed transmitter-receiver deployment, the generated channel impulse response (CIR), channel frequency response (CFR), and Radio Map results show that material binding mainly changes weak reflection, transmission, and scattering paths, reducing the number of effective paths from about 742 to about 52 while keeping the dominant path amplitude nearly unchanged.
\end{abstract}


\IEEEpeerreviewmaketitle

\section{Introduction}
\IEEEPARstart{T}{he} core objective of a digital twin is to construct a digital replica that reflects the state and behavior of a real physical object \cite{tao2019digitaltwin}. In sixth-generation (6G) network research, this concept has been extended to the joint modeling of wireless network states, propagation environments, and communication processes \cite{tao2024wirelessdt}. The IMT-2030 vision also identifies integrated sensing and communication, AI-communication integration, and related capabilities as important directions for future mobile communication systems \cite{ituM2160}. For wireless digital twins, a three-dimensional (3D) scene should not only provide visual appearance, but also support electromagnetic propagation computation.

Indoor millimeter-wave links such as 28~GHz are more sensitive to the environment. Walls, doors, windows, furniture, glass, and metal objects affect line-of-sight (LOS), reflection, transmission, and scattering paths. These interactions further change the CIR, CFR, and spatial coverage distribution \cite{rappaport2013mmwave,rappaport2019wireless}. Recent studies on 6G AI-enabled air interfaces also point out that wireless environmental information should be progressively extracted from raw sensing data into environmental features, semantics, and knowledge for channel prediction and communication optimization \cite{zhang2025aiair}.

Ray tracing is an important method for deterministic channel modeling. Platforms such as Sionna RT can compute propagation paths and output CIR, CFR, and Radio Map results when 3D geometry, antenna configurations, and material parameters are provided \cite{hoydis2023sionna}. However, existing simulation scenes often rely on manual CAD or Blender modeling. This process is time-consuming for real indoor environments and makes it difficult to fully reproduce wall thickness, door/window openings, and local blockage details.

In recent years, methods such as COLMAP, 3DGS, and SuGaR have lowered the barrier for real-scene modeling \cite{schoenberger2016sfm,kerbl2023gaussian,guedon2024sugar}. They enable camera pose recovery, high-fidelity visual representation learning, and triangular mesh extraction. The problem is that these methods are mainly optimized for visual reconstruction. A visually plausible mesh may still suffer from missing wall thickness, local broken surfaces, scale errors, and absent material information, making it difficult to use directly for millimeter-wave ray tracing.

RGB-LiDAR multimodal sensing provides a new data basis for constructing real indoor RF scenes. RGB images provide continuous viewpoints and texture information, while LiDAR point clouds provide physical scale and geometric reference. OpenScene can further provide open-vocabulary semantic cues for walls, windows, wooden furniture, metal objects, and related regions \cite{peng2023openscene}. Together, these sources support the conversion from visually observable scenes to RF-computable scenes.

This paper argues that visual reconstruction results of real indoor scenes must be further constrained by scale, regularized for RF geometry, and bound with electromagnetic materials before they can be converted into RF digital twin scenes for 28~GHz ray-tracing channel simulation. To this end, RFDT-Channel is proposed. The workflow consists of three key modules: an RGB-LiDAR 3D reconstruction and mesh extraction module, a LiDAR-guided RF geometric regularization module, and an OpenScene-assisted semantic material mapping and Sionna RT channel simulation module. The contributions are as follows.
\begin{itemize}
  \item An indoor RF digital twin scene construction workflow based on RGB-LiDAR multimodal sensing is proposed, organizing real videos, LiDAR point clouds, and visual reconstruction results into scene inputs for ray-tracing simulation.
  \item A LiDAR-guided RF geometric regularization method is designed to convert the visual mesh generated by 3DGS/SuGaR into RF-computable geometry with scale reference, wall thickness, door/window openings, and stable boundaries.
  \item OpenScene semantic segmentation results are combined with standard material models to bind electromagnetic materials such as concrete, glass, wood, and metal, and All Concrete and Multi-Material configurations are compared through CIR, CFR, and Radio Map analysis.
\end{itemize}

\section{Background and Motivation}
Millimeter-wave communication shows great potential in terms of large bandwidth and high capacity, but its propagation process is more sensitive to the environment. In indoor scenarios, LOS paths may be blocked by walls or furniture, reflection paths are affected by metal, glass, and concrete surfaces, and transmission paths are closely related to material thickness and electromagnetic parameters. Existing millimeter-wave communication studies have pointed out that high-frequency links differ significantly from traditional low-frequency cellular links in path loss, blockage sensitivity, and directional transmission \cite{rappaport2013mmwave,rappaport2019wireless}. Therefore, compared with low-frequency cellular scenarios, 28~GHz indoor links require more detailed modeling of spatial structure, transmitter-receiver positions, and material boundaries.

For a given set of propagation paths, the discrete multipath CIR can be written as
\begin{equation}
  h(\tau)=\sum_{\ell=1}^{L} a_\ell \delta(\tau-\tau_\ell),
  \label{eq:cir}
\end{equation}
where $L$ denotes the number of effective paths, and $a_\ell$ and $\tau_\ell$ denote the complex gain and propagation delay of the $\ell$-th path, respectively. This expression characterizes the distribution of LOS, reflection, scattering, and transmission paths in the delay domain. The corresponding CFR is
\begin{equation}
  H(f)=\sum_{\ell=1}^{L} a_\ell e^{-j2\pi f\tau_\ell}.
  \label{eq:cfr}
\end{equation}
This formula shows that the CFR is essentially the coherent superposition of different multipath components in the frequency domain. Once the material configuration changes path gains or the number of effective paths, the multipath distribution in the CIR and the frequency-selective fluctuations in the CFR will also change. In other words, to make millimeter-wave simulation results physically interpretable, it is not enough for the scene to have geometric boundaries; it must also carry reasonable material properties.

Visual 3D reconstruction is usually optimized for image reprojection error or rendering quality. Traditional multi-view geometry methods perform feature matching, camera pose estimation, sparse reconstruction, and multi-view stereo to obtain denser geometry. COLMAP's structure-from-motion (SfM) and multi-view stereo (MVS) pipelines are representative examples of this route \cite{schoenberger2016sfm,schoenberger2016mvs}. Active depth sensing methods emphasize depth observations and continuous surface fusion. KinectFusion uses truncated signed distance field fusion for real-time dense mapping \cite{newcombe2011kinectfusion}, while nvblox provides GPU-accelerated incremental distance field mapping for robotic scenarios \cite{millane2024nvblox}. Among learning-based scene representations, NeRF uses implicit radiance fields for high-quality novel view synthesis \cite{mildenhall2020nerf}. 3DGS represents scenes with a large number of 3D Gaussian primitives and improves training and rendering efficiency \cite{kerbl2023gaussian}. SuGaR further extracts editable meshes from Gaussian fields \cite{guedon2024sugar}. These methods are valuable for recovering visual appearance and scene structure, but RF simulation follows a different quality standard.

Stable surface normals, continuous faces, correct spatial scale, and clear physical boundaries are prerequisites for reliable ray tracing. Local holes in a visual mesh may have little impact on rendering, but they can change whether a ray penetrates a wall. Walls without thickness remove the geometric basis for transmission loss modeling. Ambiguous door and window boundaries affect diffraction and occlusion relationships. Non-manifold topology may also make ray intersection unstable. Therefore, a regularized modeling stage is necessary when moving from visual reconstruction results to an RF scene.

Recent studies also show that the key bottleneck of digital twins is shifting from whether simulation can be performed to how reliable scene assets can be obtained efficiently. For example, digital twin reconstruction for intelligent transportation systems uses onboard LiDAR point clouds and a vehicle-RSU-cloud collaborative training framework to accelerate cloud-based point-cloud compression and digital twin reconstruction \cite{chen2026dtits}. Open Wireless Digital Twin integrates OpenAirInterface with Sionna RT and uses real building data to support end-to-end wireless digital twin emulation for 5G mobility scenarios \cite{iye2025owdt}. Work on channel prediction with an uncalibrated digital twin further shows that even when electromagnetic material parameters are not fully accurate, scene geometry can still provide a useful geometric prior for combining ray tracing with a small number of measurements \cite{abouamer2025prediction}. In addition, differentiable ray-tracing research further indicates that the calibration relationship among geometric models, material properties, and measured channels significantly affects the channel-prediction capability of wireless digital twins \cite{hoydis2024learning}. In-vehicle terahertz channel research also constructs a vehicle digital twin from high-resolution point clouds and a material-property database, and imports it into Sionna for ray-tracing simulation \cite{zhu2025invehicle}. Together, these studies indicate that real sensing data, geometric reconstruction, material modeling, and ray tracing are forming an important technical route for wireless digital twins.

In indoor modeling scenarios, RGB and LiDAR are strongly complementary. RGB video provides continuous viewpoints and texture information, which helps COLMAP and 3DGS recover scene appearance. LiDAR point clouds directly describe 3D spatial structure and provide physical scale and geometric reference. In this work, a post-processing registration strategy is adopted: COLMAP, 3DGS, and SuGaR are first completed in the visual coordinate system, and then the initial mesh is scaled, spatially corrected, and regularized with the LiDAR point cloud. This avoids the complexity of front-end multi-sensor joint optimization while allowing the point cloud to improve geometric reliability during back-end RF asset construction. From the perspective of wireless environmental information, the RGB video and LiDAR point cloud used in this paper correspond to raw sensing data; the regularized geometry and material regions correspond to channel-oriented environmental features and semantic expressions; and the CIR, CFR, and Radio Map output by Sionna RT further reflect the influence of environmental information on channel results \cite{zhang2025aiair}.

Material properties in a ray-tracing scene cannot be automatically obtained from the geometric model alone. OpenScene aggregates 2D visual-language features into 3D point-cloud or mesh space, supporting open-category queries such as wall, door, window, wooden furniture, and metal cabinet. Its open-vocabulary capability can be traced to image-text contrastive learning models such as CLIP \cite{radford2021clip}, while OpenScene further transfers such visual-language features to 3D scene understanding \cite{peng2023openscene}. In RFDT-Channel, OpenScene serves as a provider of semantic and material category cues, rather than a predictor of electromagnetic parameters. Following this idea, semantic regions are mapped to materials such as concrete, glass, wood, and metal/PEC, with radio propagation parameters referenced from ITU-R P.2040 \cite{ituP2040}. Through this step, the scene is converted from visually observable to electromagnetically computable.

Compared with the above studies, the novelty of this paper does not lie in proposing a new ray-tracing algorithm or a new 3D reconstruction network. Instead, it lies in organizing real-scene multimodal sensing, visual 3D reconstruction, LiDAR geometric constraints, semantic material mapping, and Sionna RT channel generation into an executable RF digital twin scene construction workflow for 28~GHz indoor channel simulation. Specifically, this paper explicitly treats the SuGaR mesh as a visual intermediate asset rather than as the final simulation-ready model. LiDAR point clouds are then used to regularize scale, orientation, wall thickness, door/window openings, and local topology for RF computation. Finally, semantic regions provided by OpenScene are mapped to electromagnetic material configurations such as concrete, glass, wood, and metal, and the influence of material binding on channel results is evaluated through CIR, CFR, and Radio Map comparisons between All Concrete and Multi-Material. This combination distinguishes this work from wireless digital twin simulations that rely mainly on ideal CAD scenes, and also from automatic 3D reconstruction studies that focus primarily on visual appearance quality.

\section{Hardware Platform}
\subsection{Point-Cloud Acquisition Environment}
ROS~2 is used as the data communication framework for LiDAR point-cloud acquisition. ROS~2 organizes sensor data streams through nodes, topics, and message types, and allows quality-of-service policies to configure the trade-off between real-time performance, reliability, and queue depth. This node-based communication mode is suitable for separating LiDAR driving, point-cloud subscription, data saving, and later inspection into relatively independent modules. For the static indoor modeling task in this paper, ROS~2 does not participate in 3D reconstruction or channel simulation. Its role is to stably receive LiDAR output and organize point-cloud data into offline files for later processing.

During acquisition, the PandarXT-32 LiDAR publishes point-cloud data through a ROS~2 driver node, while the acquisition program subscribes to the point-cloud stream and saves the scanning results. Each point-cloud frame contains 3D coordinates, reflection intensity, timestamps, and related information. After receiving a point-cloud message, the acquisition script parses the spatial point fields and exports them as PCD/PLY files by frame or by acquisition segment.

Compared with directly saving a complete ROS~2 bag, subscribing to the point-cloud stream with a Python script provides more flexibility in saving frequency, file naming, and export format. For later Blender-based regularized modeling, point-cloud inspection, and scale correction, PCD/PLY files are also easier to load directly. Therefore, ROS~2 plays the role of point-cloud organization and storage during the multimodal data acquisition stage.

\subsection{Hardware Platform and Data Organization}
The experimental platform uses NVIDIA Jetson Orin as the edge computing unit and integrates a Hesai PandarXT-32 LiDAR and high-speed GMSL cameras. Camera videos are used for visual reconstruction, while LiDAR point clouds provide spatial geometric reference. The final recorded data include indoor RGB videos, LiDAR point-cloud frames, and global point-cloud maps in PCD/PLY format. Camera data are later extracted into RGB image sequences for COLMAP pose estimation and 3DGS training, while LiDAR point clouds are used for scale correction, spatial alignment, and RF geometric regularization. Jetson Orin provides the edge computing platform for on-site acquisition and data organization, while PandarXT-32 provides multi-line laser scanning point clouds for indoor structures.

\begin{table}[H]
\centering
\caption{Components of the multimodal acquisition platform}
\label{tab:hardware}
\begin{tabular}{m{0.32\linewidth}m{0.56\linewidth}}
\hline
Component & Function \\
\hline
NVIDIA Jetson Orin & Edge-side data acquisition and organization \\
GMSL camera & Indoor video recording and RGB image sequence acquisition \\
Hesai LiDAR & Metric-scale point-cloud acquisition and geometric reference \\
ROS~2 & Point-cloud data organization and timestamp recording \\
\hline
\end{tabular}
\end{table}

The core role of this hardware platform is to provide two complementary observations for subsequent RF scene construction: RGB image sequences required for visual reconstruction, and LiDAR point clouds required for scale correction and geometric regularization. The main hardware components and their functions are summarized in Table~\ref{tab:hardware}.

\section{Framework}
\subsection{Overall Workflow}
\begin{figure}[t]
\centering
\includegraphics[width=\linewidth]{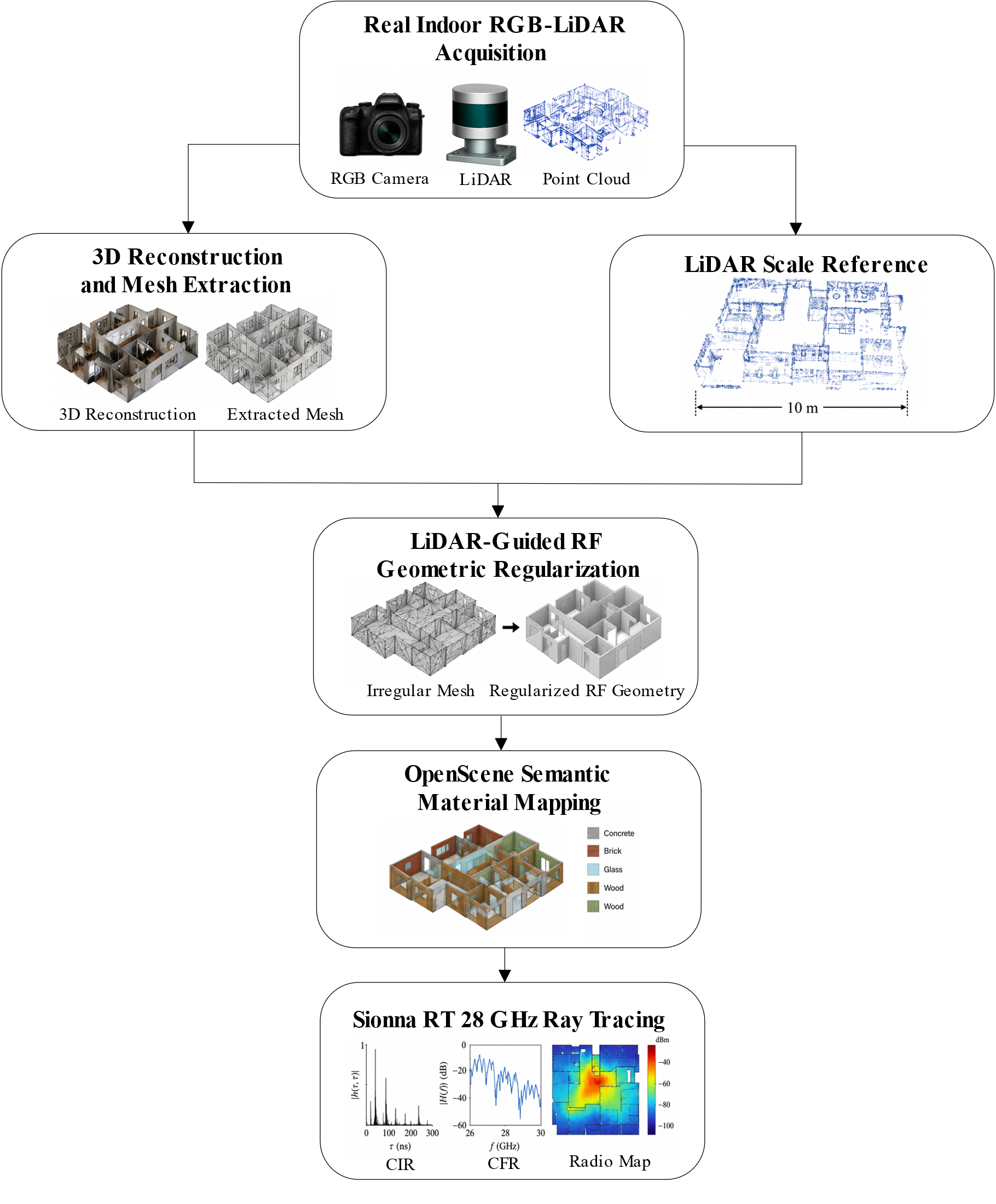}
\caption{System diagram of RFDT-Channel.}
\label{fig:workflow}
\end{figure}

The overall workflow of RFDT-Channel is shown in Fig.~\ref{fig:workflow}. It starts from a real indoor scene, where GMSL cameras and LiDAR simultaneously collect RGB videos and point-cloud data. The visual branch extracts video frames, estimates camera poses using COLMAP, trains a 3DGS representation, and applies SuGaR mesh extraction to recover an initial model with visual appearance and primary spatial structure. The point-cloud branch provides physical scale reference and a blueprint for regularized modeling. After the initial mesh is obtained, the system performs LiDAR-guided RF geometric regularization in Blender, including gravity alignment, axis normalization, wall solidification, and door/window opening construction. Then, semantic point clouds from OpenScene are used to identify material-related regions such as walls, glass, wooden structures, and metal objects, and these regions are mapped onto the regularized mesh to form geometry assets separated by material category. Finally, Sionna RT loads the RF-computable scene, binds electromagnetic materials, and outputs the CIR, CFR, and Radio Map.

Figure~\ref{fig:workflow} presents the system diagram of RFDT-Channel. Rather than emphasizing the details of a single algorithm, the figure explains how real indoor observations are sequentially converted into an RF digital twin scene usable for 28~GHz ray-tracing channel simulation.

\subsection{3D Reconstruction and Mesh Extraction}
The goal of the 3D reconstruction stage is to convert the RGB video collected in Section~III into an editable initial scene mesh. This stage does not directly generate the final RF simulation model; instead, it provides the visual and structural basis for subsequent RF geometric regularization. Specifically, the video is first extracted into an RGB image sequence at a fixed frame rate. Since the images come from the same continuous video, the viewpoint changes between frames are relatively smooth. This paper therefore uses the COLMAP sequential matching pipeline to estimate camera poses and reconstruct a sparse point cloud \cite{schoenberger2016sfm}. The camera intrinsics and extrinsics, sparse structure, and undistorted images output by COLMAP jointly form the training input for 3DGS.

The relationship between 3DGS and COLMAP can be understood as a division of labor between pose estimation and scene representation learning. COLMAP provides geometric constraints among multi-view images and determines the observation position of each image in 3D space. Under these camera pose constraints, 3DGS uses a set of 3D Gaussian primitives to represent scene appearance and spatial distribution \cite{kerbl2023gaussian}. A Gaussian primitive can be written as
\begin{equation}
  \mathcal{G}_i=\{\bm{\mu}_i,\mathbf{\Sigma}_i,o_i,\mathbf{c}_i\},
  \label{eq:gaussian_primitive}
\end{equation}
where $\bm{\mu}_i$ denotes the center of the $i$-th Gaussian, $\mathbf{\Sigma}_i$ denotes the 3D covariance matrix, $o_i$ denotes opacity, and $\mathbf{c}_i$ denotes color or spherical-harmonic color coefficients. For a spatial point $\mathbf{x}$, the response of this Gaussian is
\begin{equation}
  G_i(\mathbf{x})=\exp\left[-\frac{1}{2}(\mathbf{x}-\bm{\mu}_i)^T\mathbf{\Sigma}_i^{-1}(\mathbf{x}-\bm{\mu}_i)\right].
  \label{eq:gaussian_response}
\end{equation}
The covariance matrix not only controls the spatial influence range of a Gaussian, but also encodes local spatial orientation. To ensure that the covariance matrix is positive semi-definite, 3DGS usually parameterizes it with a rotation matrix $\mathbf{R}_i$ and a scaling matrix $\mathbf{S}_i$:
\begin{equation}
  \mathbf{\Sigma}_i=\mathbf{R}_i\mathbf{S}_i\mathbf{S}_i^T\mathbf{R}_i^T .
  \label{eq:covariance_decomposition}
\end{equation}
This means that each Gaussian is not a simple spherical point, but an anisotropic ellipsoid with orientation and scale. The direction with a larger scale corresponds to the extension of the Gaussian in that direction, while the direction with a smaller scale usually corresponds to a thinner local structure. Because Gaussian primitives carry center, scale, and orientation information, SuGaR can further extract surface geometry from the rendering representation learned by 3DGS.

During rendering, 3DGS projects 3D Gaussians onto the camera image plane. Let $\mathbf{W}$ denote the transformation from world coordinates to camera coordinates, and let $\mathbf{J}$ denote the first-order Jacobian matrix of the projection function at the current position. The 3D covariance can then be approximately projected into a 2D covariance as
\begin{equation}
  \mathbf{\Sigma}'_i=\mathbf{J}\mathbf{W}\mathbf{\Sigma}_i\mathbf{W}^T\mathbf{J}^T .
  \label{eq:projected_covariance}
\end{equation}
For a pixel $\mathbf{p}$ on the image plane, the opacity contribution of the $i$-th projected Gaussian can be written as
\begin{equation}
  \tilde{\alpha}_i(\mathbf{p})=o_i\exp\left[-\frac{1}{2}(\mathbf{p}-\mathbf{p}_i)^T(\mathbf{\Sigma}'_i)^{-1}(\mathbf{p}-\mathbf{p}_i)\right].
  \label{eq:projected_opacity}
\end{equation}
Here, $\mathbf{p}_i$ denotes the projected position of the Gaussian center on the image plane. After sorting the Gaussians from front to back by depth, the pixel color is obtained through alpha compositing:
\begin{equation}
  \mathbf{C}(\mathbf{p})=\sum_{i=1}^{N} T_i(\mathbf{p})\tilde{\alpha}_i(\mathbf{p})\mathbf{c}_i ,
  \label{eq:alpha_compositing}
\end{equation}
\begin{equation}
  T_i(\mathbf{p})=\prod_{j=1}^{i-1}\left(1-\tilde{\alpha}_j(\mathbf{p})\right).
  \label{eq:transmittance}
\end{equation}
This formula shows that the optimization objective of 3DGS mainly comes from the difference between rendered images and real images. During training, the system renders images from the Gaussian field according to the camera poses provided by COLMAP, and backpropagates the reconstruction error to optimize Gaussian centers, scales, orientations, colors, and opacities. The resulting Gaussian field has good visual fidelity, but it is still a continuous primitive set designed for image rendering and cannot directly serve as ray-intersection geometry in Sionna RT.

SuGaR converts the above 3DGS Gaussian set into an editable triangular mesh \cite{guedon2024sugar}. From the formula-level relationship, SuGaR does not start modeling from images again. Instead, it reuses the optimized $\bm{\mu}_i$, $\mathbf{\Sigma}_i$, and $o_i$ from 3DGS, and treats the Gaussian set as an implicit density field. For example, the scene density can be summarized as
\begin{equation}
  D(\mathbf{x})=\sum_i o_iG_i(\mathbf{x}).
  \label{eq:density_field}
\end{equation}
If used only for rendering, $D(\mathbf{x})$ mainly determines alpha compositing after projection. In SuGaR, however, the same density field is further interpreted as geometric evidence for scene surfaces. An intuitive way to define a surface is to regard a density level set as a potential surface:
\begin{equation}
  \mathcal{S}_{\tau}=\{\mathbf{x}\mid D(\mathbf{x})=\tau\}.
  \label{eq:level_set}
\end{equation}
The normal of this level set can be approximated by the density gradient:
\begin{equation}
  \mathbf{n}(\mathbf{x})=\frac{\nabla D(\mathbf{x})}{|\nabla D(\mathbf{x})|}.
  \label{eq:surface_normal}
\end{equation}
Therefore, the formula-level relationship between 3DGS and SuGaR can be summarized as follows: 3DGS learns a set of Gaussian primitives that can explain multi-view images through projection and alpha compositing; SuGaR reorganizes these Gaussian primitives into a density field and surface cues, encouraging better alignment between Gaussian centers, covariance principal directions, and local surfaces. In other words, 3DGS answers how to render real images with Gaussians, while SuGaR answers how to recover editable surfaces from these Gaussians.

During practical mesh extraction, SuGaR strengthens the alignment between Gaussians and surfaces, making the thin direction of a Gaussian closer to the local normal and the wide directions more consistent with the local tangent plane. It then samples points and normals near visible surfaces and obtains a triangular mesh through surface reconstruction. This forms the following relationship: COLMAP provides camera geometry, 3DGS provides a high-fidelity visual scene representation, and SuGaR converts the Gaussian representation into an editable initial mesh. This mesh preserves the main structure of the indoor scene and becomes the input for LiDAR-guided regularized modeling.

\subsection{LiDAR-Guided RF Geometric Regularization}
The initial mesh output by SuGaR is still a visual reconstruction result, not the final RF scene. Local noise, holes, broken surfaces, uneven wall boundaries, and unclear door/window contours may appear. For visual rendering, these issues may sometimes be hidden by textures or viewpoint changes. In ray tracing, however, they directly affect reflecting surface normals, occlusion boundaries, penetration path lengths, and ray-intersection stability. Therefore, RFDT-Channel treats the initial mesh as an intermediate asset that must undergo LiDAR-guided geometric regularization before entering channel simulation.

The goal of RF geometric regularization is to convert the visual mesh into ray-tracing-computable geometry, rather than to continue pursuing texture detail. The global LiDAR point cloud is used as a high-confidence spatial blueprint, and gravity alignment and axis normalization are performed in Blender so that walls, floors, and major structures remain consistent with the actual indoor space. Then, based on the relative relationship between the point cloud and the visual mesh, walls, doors, windows, furniture, and other structures are regularized.

The specific operations include using wall and floor point clouds to help determine the main room planes, applying Solidify to assign physical thickness to walls, using Boolean operations to construct door and window openings, repairing local broken surfaces and unstable boundaries, and preserving key blockage and reflection boundaries for furniture and metal objects. This process sacrifices some visual texture details, but improves the physical interpretability of geometric boundaries and the stability of ray intersection. From the perspective of communication simulation, this step completes the transition from visual reconstruction to RF-computable geometry. Here, the LiDAR point cloud provides scale and geometric reference, rather than error-free physical ground truth.

\subsection{Material Recognition and Electromagnetic Material Mapping}
A regularized mesh describes the geometric boundaries of the scene, but it does not automatically provide electromagnetic responses for different surfaces. To convert the scene from geometrically computable to electromagnetically computable, this paper uses OpenScene for open-vocabulary semantic recognition of indoor point clouds \cite{peng2023openscene}. The core idea of OpenScene is to transfer open-vocabulary features from 2D vision-language models into 3D point-cloud or mesh space, so that 3D points can obtain semantic category cues from text prompts. Compared with traditional closed-set segmentation, this is more suitable for real indoor scenes because this paper can directly query categories related to material binding, such as wall, floor, ceiling, window, wooden furniture, and metal cabinet.

In this workflow, OpenScene does not predict permittivity, conductivity, or surface roughness. Instead, it provides semantic cues for material-related regions. Specifically, this paper first sets text prompts according to material-binding requirements and separates point-cloud regions related to walls, glass structures, wooden structures, and metal objects. These semantic point clouds are then used as category masks and mapped onto the LiDAR-regularized mesh to extract the corresponding triangular faces. In this way, the point cloud provides semantic region cues, while the mesh serves as the continuous geometric carrier required for ray tracing, avoiding direct ray intersection on discrete point clouds.

\begin{figure}[t]
\centering
\includegraphics[width=\linewidth]{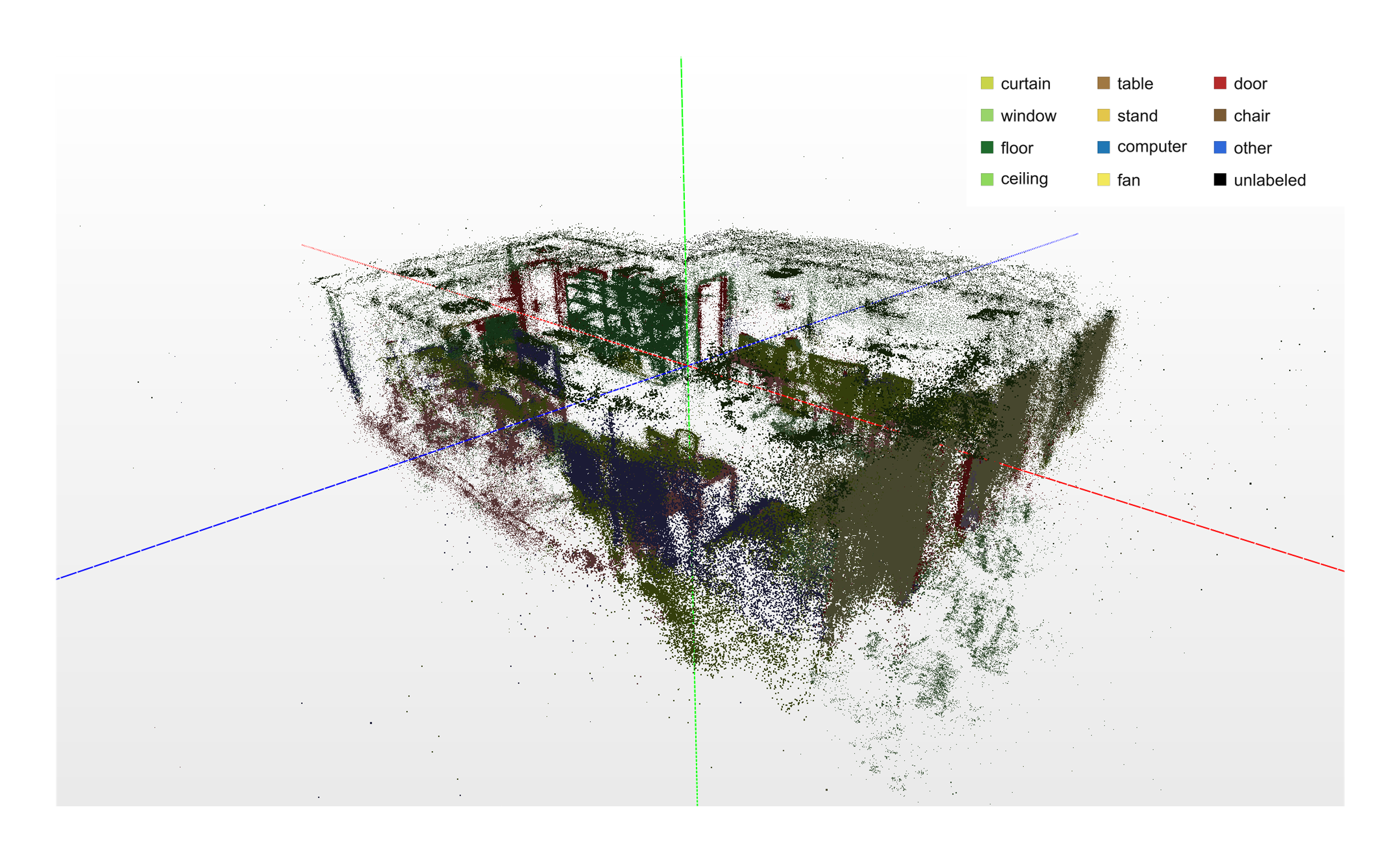}
\caption{OpenScene semantic recognition result.}
\label{fig:openscene}
\end{figure}

Figure~\ref{fig:openscene} shows an example of OpenScene semantic recognition on the indoor point cloud. Walls, floors, ceilings, doors/windows, tables/chairs, and other indoor objects are assigned different semantic colors. This result indicates that open-vocabulary 3D semantic understanding can provide cues for subsequent material-region separation. However, the categories in the figure are still visual semantic categories, and they must be further merged and checked before being mapped to electromagnetic materials such as concrete, glass, wood, and metal.

Semantic categories themselves are not equivalent to electromagnetic material parameters. This paper combines OpenScene category results with manual inspection, and then establishes mappings according to indoor object properties and the ITU-R P.2040 material model \cite{ituP2040}. Specifically, walls, floors, and ceilings are bound to concrete; windows or glass doors are bound to glass; wooden doors, tables, chairs, and wooden furniture are bound to wood; and metal cabinets and metal frames are approximated as metal.

\section{Experimental Results and Analysis}
\subsection{Material Configurations and Simulation Setup}
After geometric regularization and material separation are completed, the RF-computable scene is imported into Sionna RT. During simulation, scene description files are first generated according to the meshes of different material categories, and then loaded by Sionna RT as a ray-tracing scene. At this point, each geometry asset is no longer only a spatial boundary, but is bound to an electromagnetic material object such as concrete, glass, wood, or metal. To highlight the effect of material binding on channel results, this paper organizes two experiments, All Concrete and Multi-Material, under the same regularized mesh, transmitter-receiver positions, and antenna settings. All Concrete serves as a uniform-material baseline, while Multi-Material corresponds to the semantic electromagnetic material configuration obtained in Section~IV.

The ray-tracing tool used in this paper is Sionna RT \cite{hoydis2023sionna}. The simulation scene is named \emph{scene\_camera\_lidar}, and is derived from the 3DGS and SuGaR reconstruction result after LiDAR-guided regularization. The carrier frequency is 28~GHz, the maximum depth is set to 8, edge diffraction and scattering are enabled, and the path types include LOS, specular reflection, diffuse reflection, and refraction.

In the experiments, the transmitter and receiver positions are fixed to isolate the influence of material configuration on both a single link and the spatial coverage distribution. The transmitter is located at $(1.5,0.0,0.8)$~m, and the receiver is located at $(1.5,4.0,0.8)$~m. The horizontal distance is 4.0~m. The CFR is computed over 1024 subcarriers with a subcarrier spacing of 30~kHz, corresponding to a bandwidth of 30.72~MHz. Three types of outputs are generated by Sionna RT: CIR for delay-domain multipath analysis, CFR for frequency-selective fading analysis, and Radio Map for evaluating the spatial distribution of received signal strength.

Two material configurations are evaluated under the same regularized mesh, transmitter-receiver positions, and antenna settings. The All Concrete configuration is used as a uniform-material baseline, where all objects are assigned the same concrete material. The Multi-Material configuration assigns concrete, glass, wood, and metal according to semantic regions. Because the geometry, link distance, and antenna deployment remain unchanged, the comparison focuses on the effect of electromagnetic material assignment.

\begin{figure}[H]
\centering
\includegraphics[width=\linewidth]{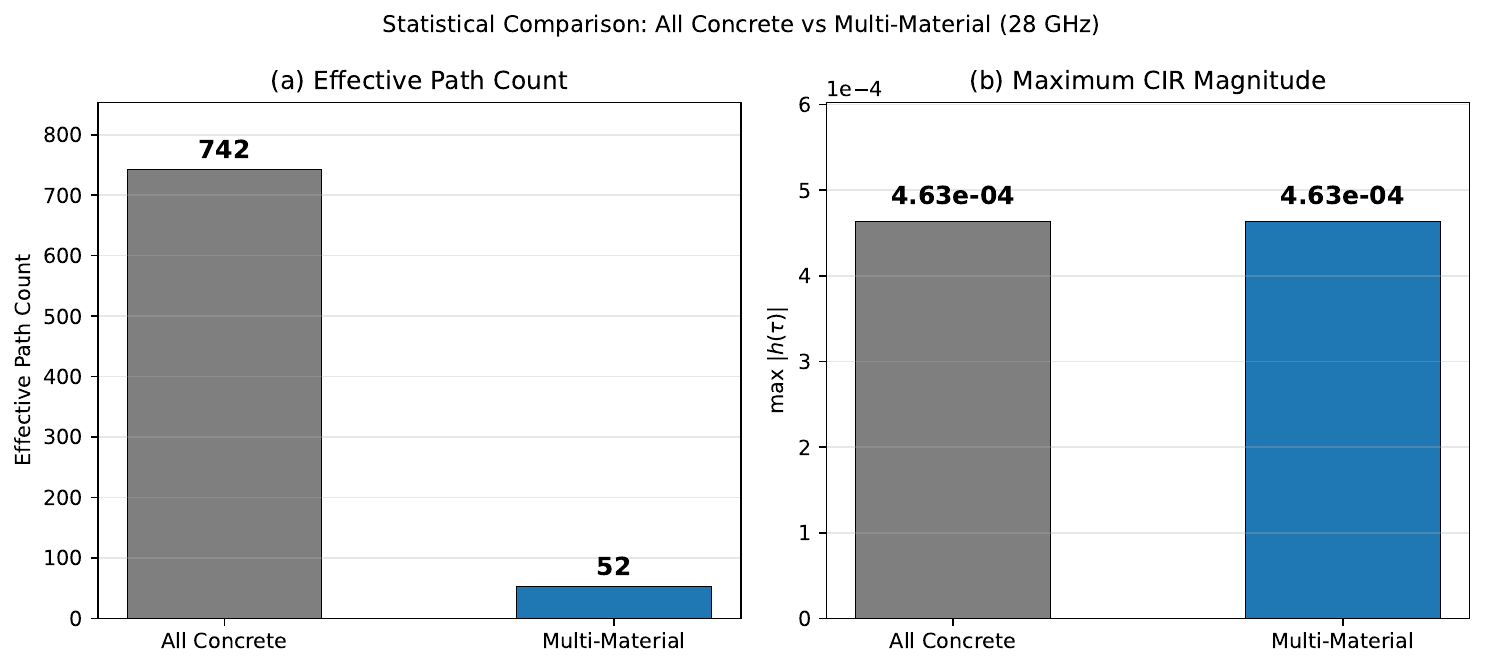}
\caption{Statistical comparison between All Concrete and Multi-Material.}
\label{fig:stats}
\end{figure}

The core statistical results are shown in Fig.~\ref{fig:stats}. The left panel compares the number of effective propagation paths retained after ray tracing, while the right panel compares the maximum CIR amplitude. These two indicators are plotted together to distinguish changes in weak multipath components from changes in the dominant path.

As shown in Fig.~\ref{fig:stats}, the number of effective paths decreases from about 742 in the All Concrete case to about 52 in the Multi-Material case, whereas the maximum CIR amplitude remains approximately $4.63\times10^{-4}$ in both cases. This indicates that semantic material binding mainly changes the retained weak reflection, transmission, and scattering paths, while the dominant path is still governed by the fixed TX/RX geometry and antenna setting in the current deployment.

\subsection{CIR and CFR Result Analysis}
\begin{figure}[H]
\centering
\includegraphics[width=\linewidth]{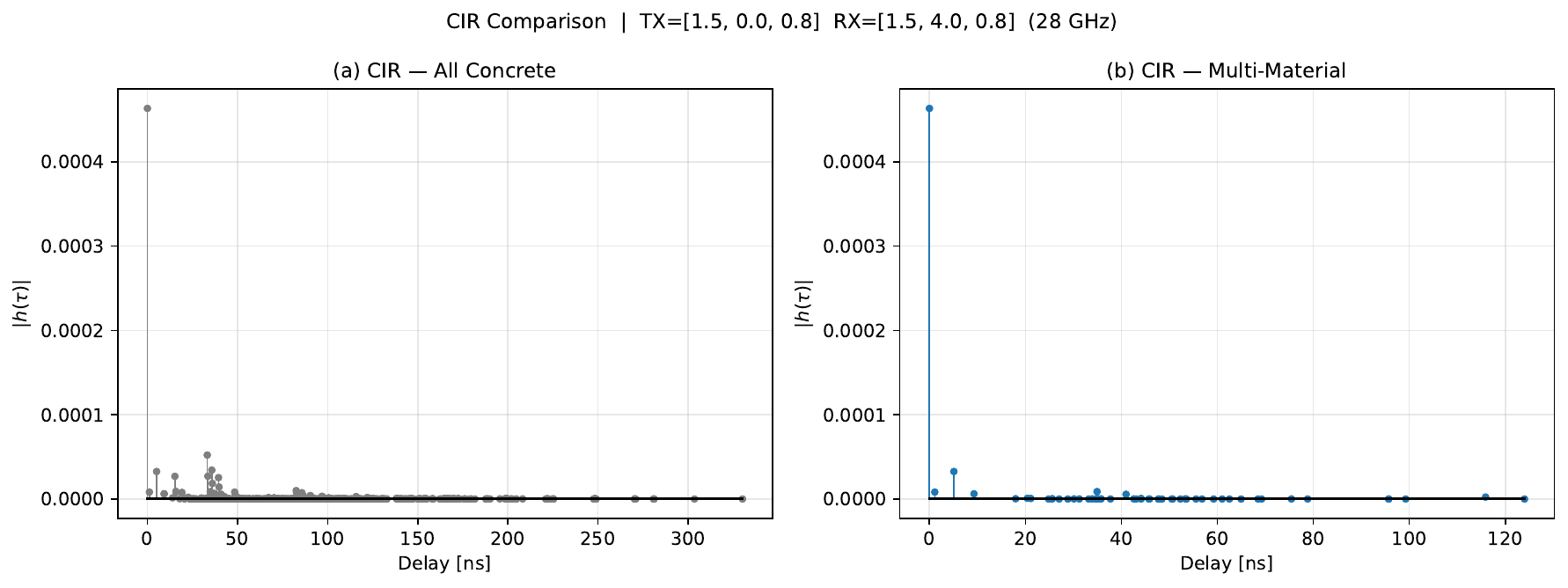}
\caption{CIR comparison between All Concrete and Multi-Material.}
\label{fig:cir}
\end{figure}

Figure~\ref{fig:cir} further shows the CIR comparison in the delay domain. The strongest peak remains at a similar amplitude in both material configurations, which is consistent with the statistical result in Fig.~\ref{fig:stats}. This is because the selected transmitter and receiver positions preserve a dominant propagation component, so the strongest path is mainly determined by the fixed TX/RX deployment, link distance, and antenna setting.

The more obvious difference lies in the weak multipath components. In the All Concrete scene, assigning the same material to all surfaces allows more weak reflection and scattering paths to remain in the delay-domain response. In the Multi-Material scene, glass, wood, metal, and concrete introduce different reflection, transmission, and attenuation effects, so some weak paths are suppressed or no longer satisfy the effective-path threshold. Therefore, material binding mainly reshapes the weak-path structure of the CIR rather than changing the dominant path amplitude.

\begin{figure}[H]
\centering
\includegraphics[width=\linewidth]{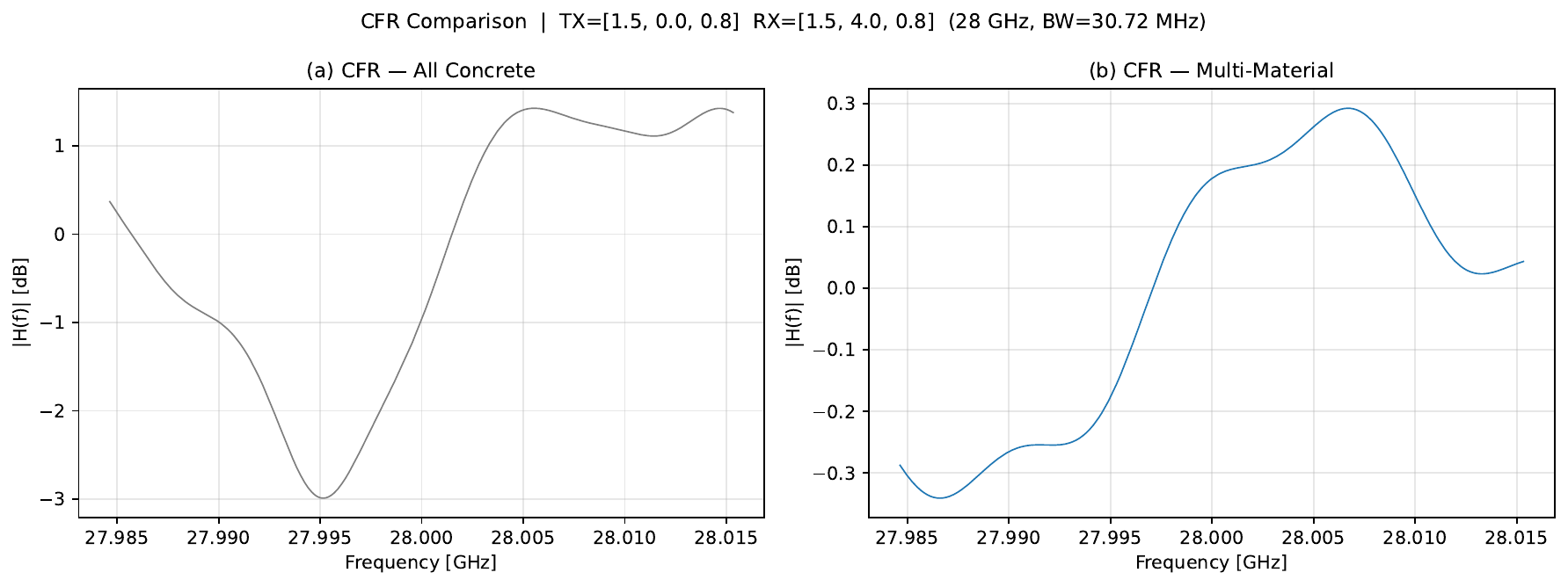}
\caption{CFR comparison between All Concrete and Multi-Material.}
\label{fig:cfr}
\end{figure}

Figure~\ref{fig:cfr} presents the corresponding CFR comparison. Since CFR is the coherent superposition of multipath components in the frequency domain, changes in the number, delay, and complex gain of paths will alter the frequency-domain fluctuation pattern.

In the All Concrete scene, the uniform material configuration retains more weak paths, so the frequency response contains richer multipath interference components. In the Multi-Material scene, different materials affect reflection, transmission, and absorption differently, changing the shape of the CFR curve. This indicates that semantic material binding influences not only the delay-domain CIR, but also the frequency-domain channel response of the fixed link.

\subsection{Radio Map Result Analysis}
\begin{figure*}[t]
\centering
\includegraphics[width=0.95\textwidth]{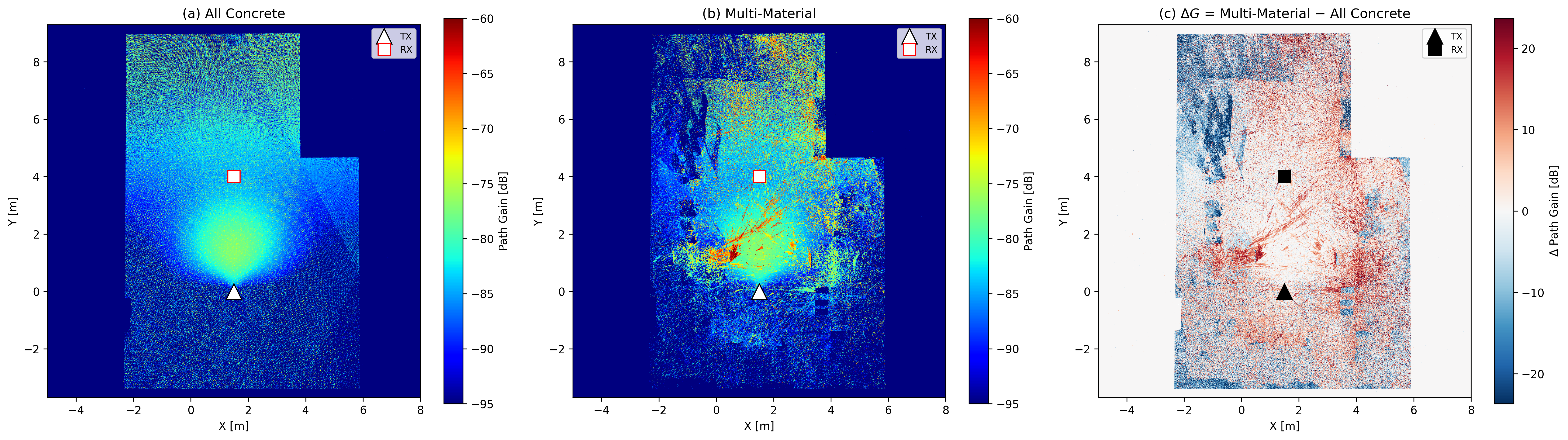}
\caption{Radio Map comparison between All Concrete and Multi-Material.}
\label{fig:radiomap}
\end{figure*}

The Radio Map characterizes the spatial coverage distribution generated by the transmitter in the indoor scene. Figure~\ref{fig:radiomap} compares the Radio Map results under the All Concrete and Multi-Material configurations and further provides their difference map. Since the two simulations share the same geometry, transmitter-receiver deployment, and antenna setting, the observed coverage variation mainly comes from material-dependent reflection, transmission, and attenuation effects.

In the All Concrete configuration, all scene objects, including walls, doors, windows, furniture, and metal objects, are assigned the same concrete material. This produces an idealized uniform-material coverage result and ignores the electromagnetic heterogeneity of real indoor environments. In contrast, the Multi-Material configuration assigns concrete, glass, wood, and metal according to semantic regions. Different materials introduce different reflection, penetration, and absorption effects, thereby changing local strong-coverage, weak-coverage, and shadowed regions in the Radio Map.

The difference map is computed as
\begin{equation}
  \Delta G=G_{\mathrm{Multi-Material}}-G_{\mathrm{All-Concrete}}.
  \label{eq:radiomap_difference}
\end{equation}
Red regions indicate locations where the received signal strength is relatively enhanced under the Multi-Material configuration, while blue regions indicate locations where it is relatively weakened. This spatial comparison shows that semantic electromagnetic material assignment affects not only the multipath structure of a fixed TX/RX link, but also the planar coverage distribution across the indoor scene.

\subsection{Discussion}
The above experiments verify the engineering feasibility of RFDT-Channel. Real RGB-LiDAR observations can pass through visual reconstruction, mesh extraction, LiDAR-guided regularization, semantic material mapping, and Sionna RT simulation to finally produce analyzable 28~GHz channel results. Compared with building a CAD scene from scratch, this workflow recovers indoor structure using real sensing data. Compared with directly using visual reconstruction meshes, it focuses on RF asset regularization and avoids treating visual appearance quality as RF simulation quality.

\section{Conclusions}
This paper presents RFDT-Channel, an RGB-LiDAR-based RF digital twin scene construction workflow for 28~GHz indoor ray-tracing channel simulation. It connects real indoor scene acquisition, COLMAP pose estimation, 3DGS high-fidelity reconstruction, SuGaR mesh extraction, LiDAR-guided regularized modeling, OpenScene semantic material mapping, and Sionna RT ray tracing into a complete pipeline. Unlike work that only pursues visual reconstruction, this paper focuses on converting 3D models into RF-computable scenes with geometric boundaries, physical scale, semantic regions, and electromagnetic material properties.

Based on \emph{scene\_camera\_lidar}, the experiments generate CIR, CFR, and Radio Map results at 28~GHz, and compare the All Concrete and Multi-Material configurations. The results show that the maximum dominant path amplitude is the same in both experiments, indicating that the current dominant LOS component is mainly governed by the fixed transmitter-receiver deployment and antenna setting. The number of effective paths decreases from about 742 to about 52, showing that semantic material binding has a clear influence on the retention of weak reflection, transmission, and scattering paths. The CFR and Radio Map differences further show that material configuration changes frequency-domain multipath superposition and spatial coverage distribution.

\bibliographystyle{IEEEtran}
\bibliography{rfdt_refs}

\end{document}